\documentclass[twocolumn,floatfix,aps,prd,showpacs]{revtex4}
\usepackage{graphicx}
\usepackage{bm}
\usepackage{epsf}

\def\agt{\mathrel{\raise.3ex\hbox{$>$}\mkern-14mu\lower0.6ex\hbox{$\sim$}}}
\def\alt{\mathrel{\raise.3ex\hbox{$<$}\mkern-14mu\lower0.6ex\hbox{$\sim$}}}

\newcommand{\beq}{\begin{equation}}
\newcommand{\eeq}{\end{equation}}
\newcommand{\beqn}{\begin{eqnarray}}
\newcommand{\eeqn}{\end{eqnarray}}
\newcommand{\pa}{\partial}

\newcommand{\varep}{\varepsilon}

\newcommand{\br}{{\mbox{\boldmath$r$}}}

\begin{document}

\title{Merger of black hole-neutron star binaries: nonspinning black hole case}

\author{Masaru Shibata$^1$}

\author{Koji Ury\=u$^2$}

\affiliation{$^1$Graduate School of Arts and Sciences, 
University of Tokyo, Komaba, Meguro, Tokyo 153-8902, Japan\\
$^2$ Department of Physics, University of Wisconsin-Milwaukee, 
P.O. Box413, Milwaukee, WI 53201, USA}

\begin{abstract}
We perform a simulation for merger of a black hole (BH)-neutron star
(NS) binary in full general relativity preparing a quasicircular state
as initial condition. The BH is modeled by a moving puncture with no
spin and the NS by the $\Gamma$-law equation of state with $\Gamma=2$. 
Corotating velocity field is assumed for the NS.  The mass of the
BH and the rest-mass of the NS are chosen to be $\approx 3.2
M_{\odot}$ and $\approx 1.4 M_{\odot}$ with relatively large radius of
the NS $\approx 14$ km. The NS is tidally disrupted near the innermost 
stable orbit but $\sim 80\%$ of the material is swallowed into the
BH with small disk mass $\sim 0.3M_{\odot}$ even for such small BH
mass $\sim 3M_{\odot}$.  The result indicates that the system of a BH
and a massive disk of $\sim M_{\odot}$ is not formed
from nonspinning BH-NS binaries, although a disk of mass $\sim 0.1M_{\odot}$
is a possible outcome. 
\end{abstract}

\maketitle

Merger of black hole (BH)-neutron star (NS) binaries is one of likely
sources of kilo-meter size laserinterferometric gravitational wave
detectors.  Although such system has not been observed yet,
statistical studies based on the stellar evolution synthesis suggest
that the merger will happen more than 10\% as frequently as the merger
of binary NSs \cite{grb1,BHNS}. Thus, the detection of such system
will be achieved by laserinterferometers in near future. 

According to a study based on the tidal approximation (which is 
referred to as a study for configuration of a Newtonian star in
circular orbits around a BH in its relativistic tidal field; e.g.,
\cite{F,shibata,WL,ISM,FBSTR}), the fate is classified into two cases,
depending on the mass ratio $q \equiv M_{\rm NS}/M_{\rm BH}$, where
$M_{\rm BH}$ and $M_{\rm NS}$ denote the masses of BH and NS,
respectively. For $q \alt q_c$, the NS of radius $R$ will be swallowed
into the BH horizon without tidal disruption before the orbit reaches
the innermost stable circular orbit (ISCO) \cite{WL,ISM}, while for $q
\agt q_c$, NS may be tidally disrupted before plunging into BH. Here,
the critical value of $q_c$ depends on the BH spin and equation of
state (EOS) of NS, and for the nonspinning case with stiff EOSs, $q_c
\approx 0.3$--$0.35(R/5M_{\rm NS})^{-3/2}$ \cite{ISM}. (Throughout
this paper, we adopt the geometrical units $c=G=1$.)

The second case has been studied with great interest because of the
following reasons. (i) Gravitational waves at tidal disruption will
bring information about the NS radius since the tidal disruption limit
depends sensitively on it \cite{valli}. The relation between the mass
and the radius of NSs may be used for determining the 
EOS of high density matter \cite{lindblom}. (ii) Tidally disrupted
NSs may form a massive disk of mass $\sim 0.1$--$1M_{\odot}$ around the BH
if the tidal disruption occurs outside the ISCO. Systems consisting of
a BH and a massive, hot disk have been proposed as one of likely
sources for the central engine of gamma-ray bursts (GRBs) with a 
short duration \cite{grb2}, and hence, merger of low-mass BH and
NS is a candidate. 

However, the scenario based on the tidal approximation studies may be
incorrect since gravitational radiation reaction and gravitational
effects of NS to the orbital motion are ignored. Radiation reaction
shortens the time available for tidally disrupting NSs. The gravity of
NS could increase the orbital radius of the ISCO and hence the
critical value of the tidal disruption, $q_c$, may be larger in reality. 
Miller \cite{CMiller} estimates these ignored effects and suggests
that NSs of canonical mass and radius will be swallowed into BH
without tidal disruption. Moreover, NSs are described by the Newtonian
gravity in the tidal approximation. If we treat it in general
relativity, the gravity is stronger and hence tidal disruption is less
likely.

Tidal disruption of NSs by a BH has been investigated in the Newtonian
\cite{Newton} and approximately general relativistic (GR) simulation
\cite{FBST,FBSTR}. However, a simulation in full general relativity is
required (see \cite{loffler} for an effort). In this paper, we present
our first results for fully GR simulation, performed by our new code
which has been improved from previous one \cite{STU0,STU}; we enable
to handle orbiting BHs adopting the moving puncture method recently
developed \cite{BB2} (see also \cite{BB4} for detailed calibration).
As the initial condition, we prepare a quasicircular state computed in
a new formalism described below.  In this paper, we focus on whether
NSs of realistic mass and radius is tidally disrupted to form a
massive disk around nonspinning BHs.  We will illustrate that a disk
with mass $\sim M_{\odot}$ is an unlikely outcome for plausible values
of NS mass and radius although a disk of mass of $O(0.1M_{\odot})$ is
possible.

\noindent
\underline{\em{Formalism for a quasicircular state}} 

Three groups have worked in computing quasicircular states of BH-NS
binaries \cite{Miller,TBFS,GRAN}. Here, we propose a new method for
computing accurate quasicircular states that can be used for numerical
simulation in the moving puncture framework \cite{BB2}. 

Even just before the merger, it is acceptable to assume that BH-NS
binaries are in a quasicircular orbit since the time scale of
gravitational radiation reaction is a few times longer than the
orbital period. Thus, we assume the presence of a helical Killing
vector around the mass center of the system, $\ell^{\mu}=(\pa_t)^{\mu}
+\Omega (\pa_{\varphi})^{\mu}$, 
where the orbital angular velocity $\Omega$ is constant.  

Next, we assume that the NS is corotating around the mass center of
the system. Irrotational velocity field is believed to be more
realistic for BH-NS binaries \cite{KBC} but we do not choose it in
this paper.  The NSs in corotation are more subject to
tidal disruption than in irrotation (e.g., \cite{shibata}) since the
outer part has larger angular momentum than the inner part and thus
outward mass ejection more easily occurs. If a corotating NS is stable
against tidal disruption, this will be also the case for the
irrotational NS of the same mass and radius.

The assumption of corotating velocity field in the helical symmetric
spacetime yields the first integral of the Euler equation, 
$
h^{-1} u^t  ={\rm const},\label{feuler}
$
where $h$ is specific enthalpy defined by $1+\varepsilon+P/\rho$,
and $\varep$, $P$, and $\rho$ are specific internal energy, pressure,
and rest-mass density, respectively. In the present work, we adopt the
$\Gamma$-law EOS with $\Gamma=2$; $P=\rho\varep=\kappa \rho^2$
with $\kappa$ an adiabatic constant. $u^{\mu}$ denotes the four
velocity and $u^t$ its time component. Assumption of corotation
implies $u^{\mu}=u^t \ell^{\mu}$. 

For a solution of geometric variables of quasicircular orbits, we
adopt the conformal flatness formalism for three-geometry. In this
formalism, the solution is obtained by solving Hamiltonian and
momentum constraint equations, and an equation for the time slicing
condition which is derived from $K_k^{~k}=0$ where $K_{ij}$ is the
extrinsic curvature and $K_k^{~k}$ its trace \cite{IWM}. 
Using the conformal factor $\psi$ and the rescaled tracefree
extrinsic curvature $\hat A_{i}^{~j}=\psi^6 K_{i}^{~j}$, 
these equations are respectively written 
\beqn
&& \Delta \psi = -2\pi \rho_{\rm H} \psi^5 -{1 \over 8}
\hat A_{i}^{~j} \hat A_{j}^{~i}\psi^{-7}, \label{ham2} \\
&& \hat A^{~j}_{i~,j} = 8\pi J_i \psi^6,\label{mom2}
\\
&& \Delta \Phi = 2\pi \Phi \Big[\psi^4 (\rho_{\rm H} + 2 S)
+{7 \over 16\pi} \psi^{-8}\hat A_{i}^{~j} \hat A_{j}^{~i}\Big],
\label{alpsi}
\eeqn
where $\Delta$ denotes the flat Laplacian, 
$\rho_{\rm H}=\rho h (\alpha u^t)^2-P$, $J_i=\rho h u_i$
and $S=\rho h [(\alpha u^t)^2-1]+3P$. 
Here, $\alpha$ is the lapse function and 
$\Phi$ is defined by $\Phi\equiv \alpha\psi$. 

We solve these equations in the framework of the puncture BH
\cite{BB,BB2,hannam}. Assuming that the puncture is located at $\br_{\rm P}$, 
we set $\psi$ and $\Phi$
\beqn
\psi=1+{M_{\rm P} \over 2 r_{\rm BH}} + \phi
\ \ \mbox{and} \ \  
\Phi=1 - {C \over r_{\rm BH}} + \eta,
\eeqn
where $M_{\rm P}$ and $C$ are positive constants, 
and $r_{\rm BH}=|x^k_{\rm BH}|$ ($x^k_{\rm BH}=x^k-x^k_{\rm P}$).
Then elliptic equations for $\phi$ and $\eta$ are derived.
The constant $M_{\rm P}$ is arbitrarily given, while 
$C$ is determined from the virial relation (e.g., \cite{vir})
\beqn
\oint_{r \rightarrow \infty} \pa_i \Phi dS^i=
-\oint_{r \rightarrow \infty} \pa_i \psi dS^i=2\pi M, 
\eeqn
where $M$ is the ADM mass. The mass center is determined from the
condition that the dipole part of $\psi$ at spatial infinity is
zero.


Equation (\ref{mom2}) is rewritten setting 
\beqn
\hat A_{ij}(=\hat A_i^{~k}\delta_{jk})
=W_{i,j}+W_{j,i}-{2 \over 3}\delta_{ij} \delta^{kl}
W_{k,l}+K^{\rm P}_{ij},\label{hataij}
\eeqn
where $K^{\rm P}_{ij}$ denotes the weighted
extrinsic curvature associated with linear momentum of a puncture BH; 
\beqn
K^{\rm P}_{ij}={3 \over 2 r_{\rm BH}^2}\biggl(n_i P_j +n_j P_i
+(n_i n_j-\delta_{ij}) P_k n_k \biggr). 
\eeqn
Here, $n^k=n_k=x^k_{\rm BH}/r_{\rm BH}$. $P_i$ denotes linear momentum of
the BH, determined from the 
condition that the total linear momentum of system should be zero; 
\beqn
P_i=-\int J_i \psi^6 d^3x. \label{P}
\eeqn
The RHS of Eq. (\ref{P}) denotes the total linear momentum
of the companion NS.
Then, the total angular momentum of the system is derived from
\beqn
J=\int J_{\varphi} \psi^6 d^3x + \epsilon_{zjk} r_{\rm P}^j \delta^{kl}P_l. 
\eeqn

The elliptic equation for $W_i(=W^i)$ is 
\beqn
\Delta W_i + {1 \over 3}\pa_i \pa_k W^k=8\pi J_i \psi^6. \label{weq}
\eeqn
Denoting $W_i=7 B_i - (\chi_{,i}+B_{k,i} x^k)$ where $\chi$ and $B_i$
are auxiliary functions \cite{gw3p2}, Eq. (\ref{weq}) is decomposed as 
\beq
\Delta B_i = \pi J_i \psi^6,~~~\Delta \chi= -\pi J_i x^i \psi^6.
\eeq

Computing BH-NS binaries in a quasicircular orbit requires to
determine the shift vector even in the puncture framework. This is
because $u_i$ has to be obtained [it is derived from $u_k=\delta_{ki}
u^t \psi^4 (v^i + \beta^i)$ where $v^i=\Omega \varphi^i$].  
The relation between 
$\beta^i$ and $\hat A_{ij}$ is written 
\beqn
\delta_{jk} \pa_i \beta^k +\delta_{ik} \pa_j \beta^k
-{2\over 3}\delta_{ij} \pa_k \beta^k ={2\alpha \over \psi^6} \hat A_{ij}.
\eeqn
Operating $\delta^{jl}\pa_l$, an elliptic equation is derived 
\beqn
\Delta \beta^i + {1 \over 3} \delta^{ik} \pa_k\pa_j\beta^j
=2 \pa_j (\alpha \psi^{-6}) \hat A^{ij}+16\pi\alpha J_j \delta^{ij},
\label{betaeq}
\eeqn
which is solved in the same manner as that for $W_i$. 


We have computed several models of quasicircular states and found
that the relation between $\Omega$ and $J$
approximately agrees with the 3rd Post-Newtonian relation \cite{Luc}.
This makes us confirm that this approach is a fair way for preparing 
quasicircular states.  We also found that in this method, the shift 
vector at $\br=\br_{\rm P}$ automatically satisfies the condition
$\beta^{\varphi}=-\Omega$ within the error of a few \%. This implies
that the puncture is approximately guaranteed to be in a corotating
orbit in the solution. 

\noindent
\underline{{\em Numerical results for quasicircular states}}

\begin{table}[tb]
\begin{center}
\caption{Parameters of quasicircular states.  Mass parameter of puncture, mass
of BH, rest-mass of NS, mass and radius of NS in isolation,
total mass of the system, nondimension angular momentum parameter,
orbital period in units of $M$, and compactness of the system defined
by $C_o=(M\Omega)^{2/3}$.  Mass of BH is computed from the area of the
apparent horizon $A$ as $(A/16\pi)^{1/2}$. Mass is shown
in units of $M_{\odot}$.
}
\begin{tabular}{cccccccccc} \hline
 & $M_{\rm P}$ & $M_{\rm BH}$ & $M_{*}$ & $M_{0\rm NS}$ 
& $R$ (km)  & $M$ & $J/M^2$ & $P_0/M$ & $C_o$ \\ \hline
A & 3.13 & 3.21 & 1.40 & 1.30 & 13.8 & 4.47 & 0.729 & 119 & 0.141 \\ 
B & 3.13 & 3.21 & 1.40 & 1.30 & 13.8 & 4.47 & 0.720 & 110 & 0.150 \\  \hline 
\end{tabular}
\end{center}
\end{table}

\begin{figure*}[t]
\begin{center}
\epsfxsize=2.25in
\leavevmode
\epsffile{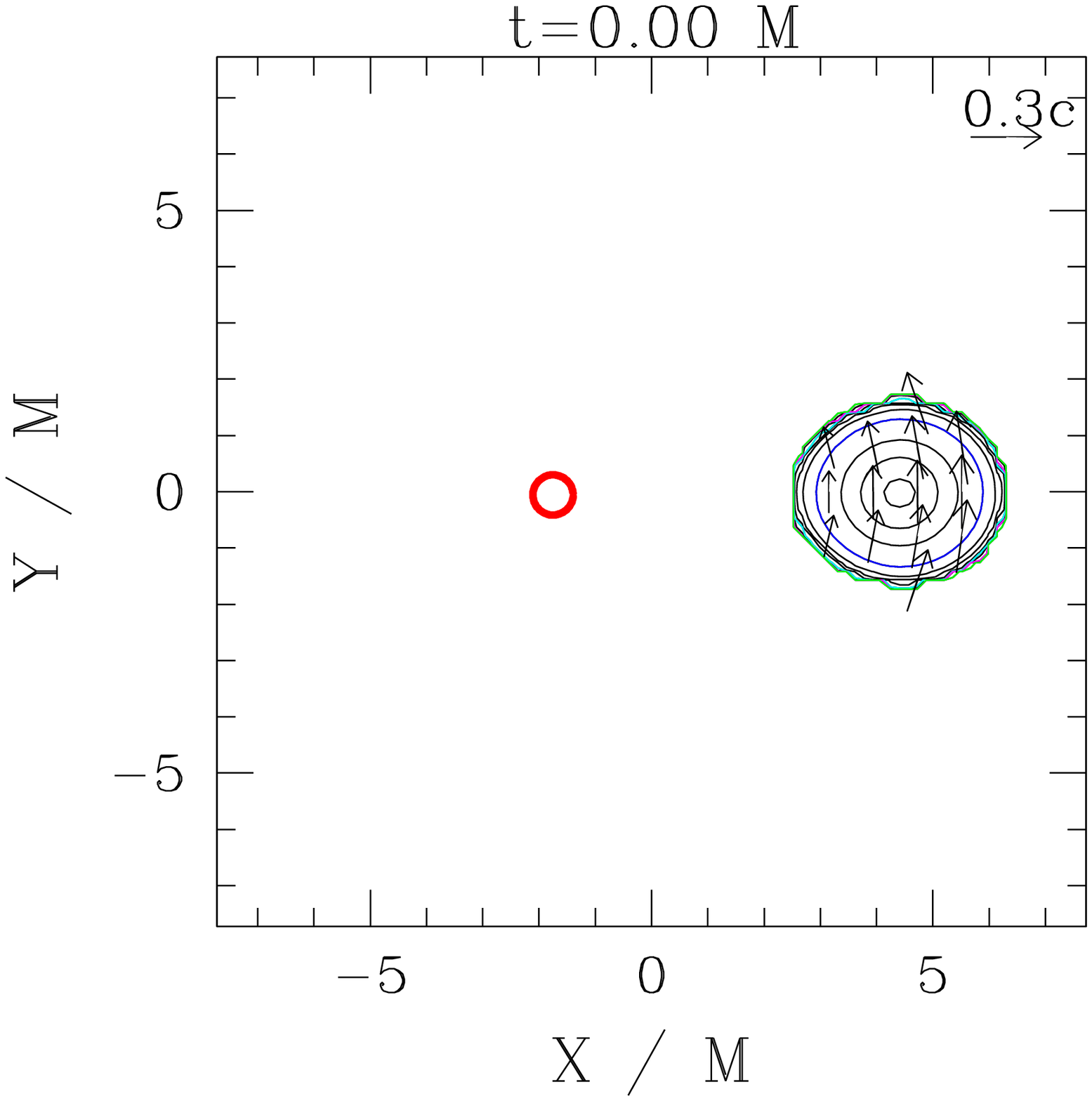}
\epsfxsize=2.25in
\leavevmode
\hspace{-1.75cm}\epsffile{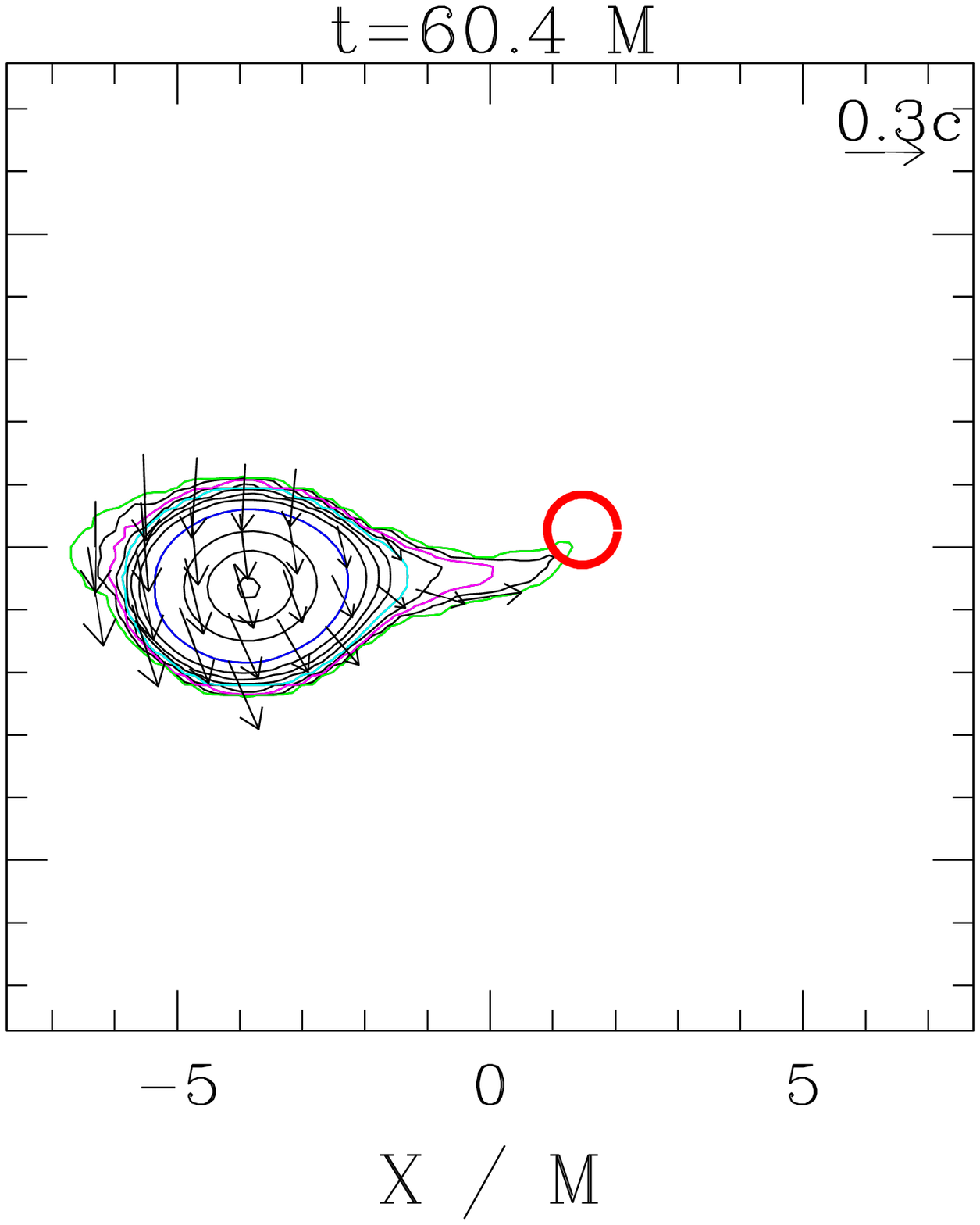} 
\epsfxsize=2.25in
\leavevmode
\hspace{-1.75cm}\epsffile{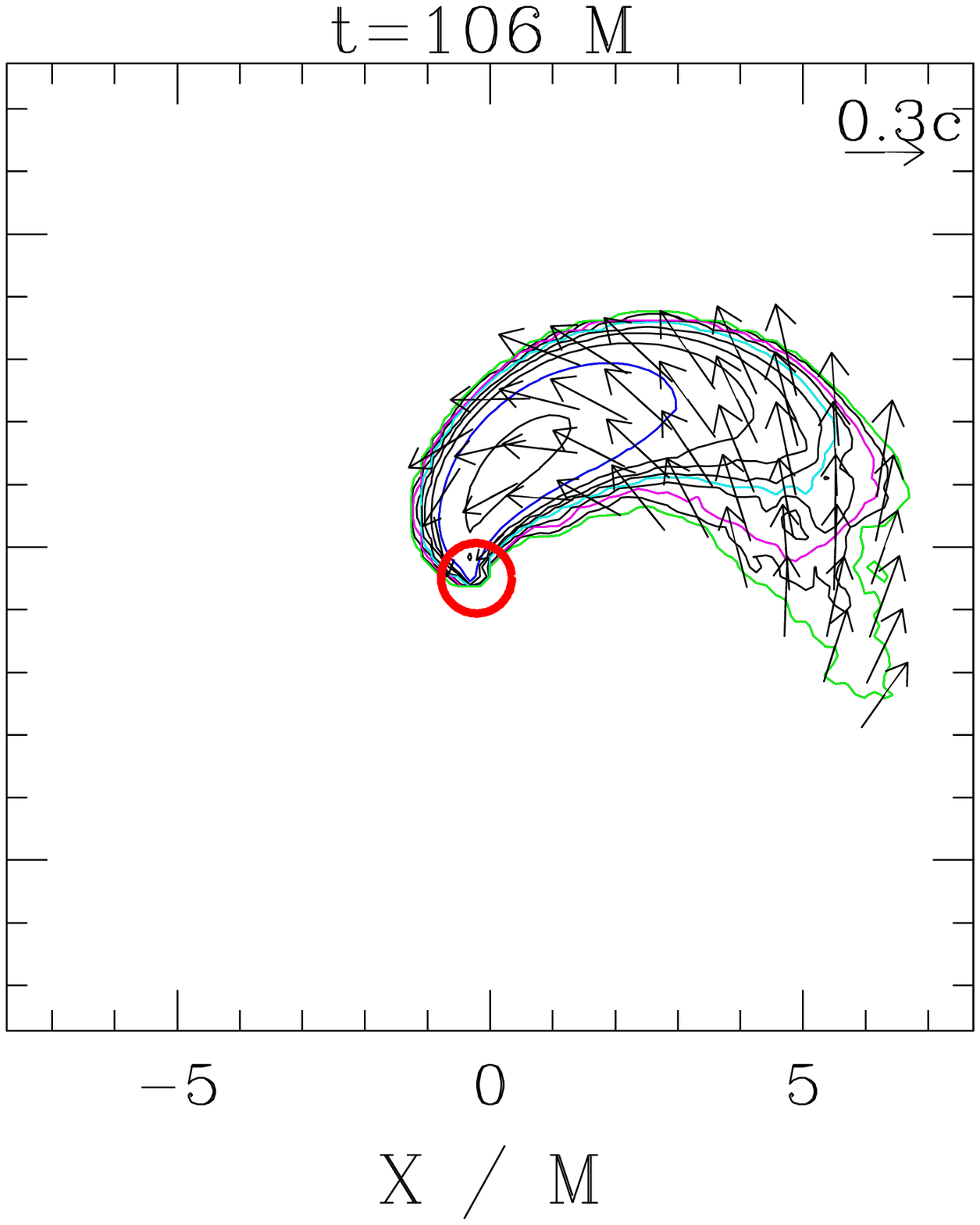} 
\epsfxsize=2.25in
\leavevmode
\hspace{-1.75cm}\epsffile{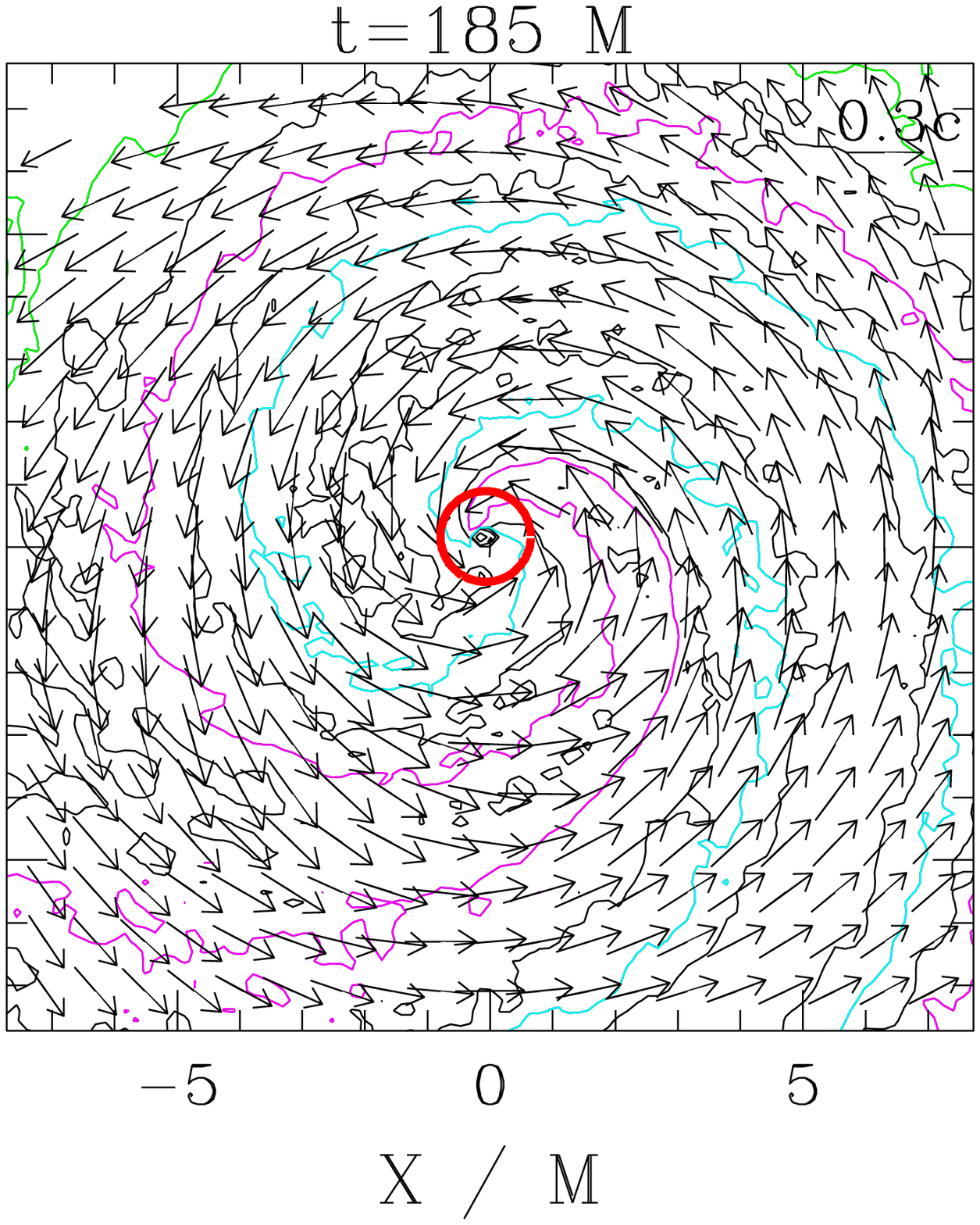}
\vspace{-8mm}
\caption{Snapshots of the density contour curves for $\rho$ in the
equatorial plane for model A.  The solid contour curves are drawn for
$\rho= (1+2i)\times 10^{14}~{\rm g/cm^3}~(i=1, 2, 3)$ and
for $10^{14-0.5 i}~{\rm g/cm^3}~(i=1 \sim 8)$. 
The maximum density at $t=0$ is $\approx 7.2 \times 10^{14}~{\rm g/cm^3}$. 
The blue, cyan, magenta, and green curves denote
$10^{14}$, $10^{12}$, $10^{11}$, and $10^{10}~{\rm g/cm^3}$, respectively.
Vectors indicate the velocity field $(v^x,v^y)$, and the scale is shown in the
upper right-hand corner. The thick (red) circles are apparent horizons.
\label{FIG1}}
\end{center}
\end{figure*}

BH-NS binaries in quasicircular orbits have been computed for a wide 
variety of models with $q=M_{*}/M_{\rm BH} \approx 0.3$--0.5 where
$M_{*}$ denotes baryon rest-mass of the NS. In the present work, the 
compactness of spherical NSs with rest-mass $M_{*}$ is chosen to be
$\approx 0.14$. In the $\Gamma$-law EOSs, the mass and radius of NS 
are rescaled by changing the value of $\kappa$: In the following we
fix the unit by setting that $M_{*}=1.4M_{\odot}$. In this case, $R
\approx 13.8$ km, the gravitational mass is $\approx 1.30 M_{\odot}$,
and $M_{*}/\kappa^{1/2}=0.147$. 
According to theories for NSs based on realistic nuclear
EOSs \cite{EOS}, the radius of NS of $M_{\rm NS} \approx 1.4 M_{\odot}$
is 11--13 km. Thus the radius chosen here is slightly larger than that
of realistic NSs and is more subject to tidal disruption. 

In computation, we focus only on the orbit of slightly outside of
ISCO. Table I shows the quantities for selected quasicircular
states with $q \approx 0.4$; model A is that used for the following
numerical simulation and model B is very close to the tidal
dispruption limit of approximately the same mass as that of model A,
showing that the model A has an orbit slightly outside the tidal
disruption limit.

The tidal approximation studies suggest that for $q \agt q_* \equiv
0.35(R/5M_{\rm NS})^{-3/2}[(M_{\rm BH}\Omega)^{-1}/6^{3/2}]$, NSs with
$\Gamma=2$ could be tidally disrupted by a nonspinning BH \cite{ISM}.
Here, $\Omega=M_{\rm BH}^{-1}/6^{3/2}$ is the angular velocity of the
ISCO around nonspinning BHs.  For model B, $q_* \approx 0.32$, and
hence, $q > q_*$.  According to the tidal approximation studies
\cite{WL,ISM}, such NS should be unstable against tidal disruption.
Nevertheless, such equilibrium exists, proving that the tidal
disruption limit in the framework of the tidal approximation does not
give correct answer. Our studies indicate that the critical value
$q_*$ is $\approx 0.43(R/5M_{\rm NS})^{-3/2}[(M_{\rm BH}\Omega)^{-1}
/6^{3/2}]$; tidal disruption of NS is much less likely than in the
prediction by the tidal approximation \cite{WL,ISM}. For the typical
NS of radius $R \sim 5M_{\rm NS}$ and mass $M_{\rm NS} \sim
1.4M_{\odot}$, $M_{\rm BH} \alt 3.3M_{\odot}$ will be necessary for
$(M_{\rm BH}\Omega)^{-2/3}\geq 6$; this implies that canonical NSs
will not be tidally disrupted outside ISCO by most of nonspinning BHs
of mass larger than $\sim 3M_{\odot}$. Tidal disruption occurs only
for NSs of relatively large radius and only for orbits very close to
ISCO.

\noindent
\underline{{\em Simulation for merger}}

Even if tidal disruption of an NS occurs near ISCO, a massive disk may
be formed around BH. To investigate this possibility we perform a
numerical simulation adopting model A.

For the simulation, we initially reset the lapse (i.e., $\Phi$)
since the relation $\alpha \geq 0$ should hold. In
the present approach we give $\Phi$ at $t=0$ by the following equation 
\beqn
\Phi &= & \eta +{1 + 0.1 X^4 \over 1+\sum_{m=1}^{3} X^m +1.1X^4}~,  
\eeqn
where $X=C/r_{\rm BH}$. 
Then, $\alpha=0$ only at puncture and otherwise $\alpha >0$.
Furthermore, for $r_{\rm BH} >C$, the values of $\Phi$ quickly
approch to those of the quasicircular states. 

The numerical code for hydrodynamics is the same as that for
performing merger of NS-NS binaries (high-resolution central scheme)
\cite{STU}. However, we change equations for $\alpha$, $\beta^i$, and
$\psi$, and numerical scheme of handling the transport terms of
evolution equations for geometries. For $\alpha$ and $\beta^i$ we solve
\beqn
&&(\pa_t -\beta^i \pa_i)\ln \alpha = -2 K_k^{~k}, 
\label{lapse} \\
&&\pa_t \beta^i=0.75 \tilde \gamma^{ij} (F_j +\Delta t \pa_t F_j),
\label{shift} 
\eeqn
where $\tilde \gamma_{ij}$ is the conformal three-metric and
$F_i=\delta^{jk}\pa_j \tilde \gamma_{ik}$. $\Delta t$ denotes the
time step for the simulation and the second term in the RHS
of Eq. (\ref{shift}) is introduced for stabilization. 
The equation for the conformal factor is also changed to
\beqn
\pa_t \psi^{-6} - \pa_i(\psi^{-6}\beta^i)= (\alpha K_k^{~k}
-2\pa_i\beta^i)\psi^{-6}, 
\eeqn
since $\psi$ diverges at the puncture \cite{BB2}. 

In addition, we have improved numerical scheme for the transport term of
geometric variables $(\pa_t - \beta^i\pa_i )Q$ where $Q$ is one of the
geometric variables: First, we rewrite this term to $\pa_t Q-\pa_i
(Q\beta^i)+Q \pa_i\beta^i$ and then apply the same scheme as in
computing the transport term of the hydrodynamic equations to the
second term (3rd-order interpolation scheme \cite{STU0}). We have 
found that for evolving punctures, such high-resolution scheme for the
transport term in the geometric variables are crucial. This is
probably because of the fact that near punctures, some of geoemetric
variables steeply vary and so is the term $\beta^i\pa_i Q$. 
For other terms in the Einstein's equation, we use the
2nd-order finite differencing as in \cite{STU0,STU}.  (Note that in
the case of nonuniform grid, 4-point finite differencing is adopted
for $Q_{,ii}$ since 3-point one is 1st-order.) After we performed 
most of runs, we repeated some of computations with 
a higher-order scheme as used in \cite{BB2,BB4}.  With such scheme, 
convergent results are obtained with a relatively large grid spacing.
However, the results are qualitatively unchanged and the extrapolated
results (which are obtained in the limit of zero grid spacing)
are approximately identical. 

We adopt the cell-centered Cartesian, $(x, y, z)$, grid to avoid the 
situation that the location of punctures (which always stay in the 
$z=0$ plane) coincides with the grid location. The equatorial plane 
symmetry is assumed; the grid size is $(2N, 2N, N)$ for $x$-$y$-$z$. 
Following \cite{SN}, we adopt a nonuniform grid; in the present 
approach, a domain of $(2N_0, 2N_0, N_0)$ grid zone is covered with 
a uniform grid of the spacing $\Delta x$ and outside the domain, the
grid spacing is increased according to $\xi\tanh[(i-N_0)/\Delta
i]\Delta x$ where $i$ denotes the $i$-th grid point in each direction. 
$N_0$, $\Delta i$, and $\xi$ are constants.  $\Delta x/M_{\rm P}$
is chosen to be $1/8$, $9/80$, $1/10$, 7/80, and 3/40.
As shown in \cite{BB2}, such grid spacing can resolve moving punctures.  
For $(N, N_0, \Delta i, \xi, \Delta x/M_p)$, we choose 
(i) (160,105,30,4.5,1/8), (ii) (200,105,30,4.5,1/8),
(iii) (200,105,30,5,1/8), (iv) (220,125,30,5,9/80),
(v) (220,125,30,6,1/10), (vi) (220,140,30,7,7/80), and (vii) (220,150,9,3/40).
For $\Delta x=M_{\rm P}/8$ and $N=160$,
we chose other values of $N_0$ and $\Delta i$, and found that 
results depend weakly on them. 
The grid covers a cube of edge length $2L$: For (i)--(vii), 
$L/\lambda=0.46$, 0.78, 0.83, 0.78, 0.78, 0.65, and 0.59,
respectively, where $\lambda$ is the wavelength of gravitational waves
at $t=0$. 

For a test, we performed a simulation for merger of two nonspinning
BHs adopting the same initial condition as used in \cite{BB2}.  We
focused particularly on the merger time and found that it varies with
improving grid resolution. By the extrapolation, a true merger time is
estimated. It is found that our result is $\approx 19M$ and agrees
with those of \cite{BB2} (see \cite{SU000} for our results).
This indicates that our code can follow moving punctures as in \cite{BB2}. 

Figure 1 shows evolution of contour curves for $\rho$ and velocity
vectors for $v^i$ in the equatorial plane together with the location
of apparent horizons at selected time slices for (vii). Due to
gravitational radiation reaction, the orbital radius decreases and
then the NS is elongated (2nd panel). Because of the elongation, the
quadrupole moment of the NS is amplified and the attractive force
between two objects is strengthen \cite{LRS}. This effect accelerates
an inward motion and, consequently, the NS starts plunging to the BH
at $t \sim 90M$. Soon after this time, the NS is tidally disrupted;
but the tidal disruption occurs near the ISCO and hence the material
in the inner part is quickly swallowed into the BH (3rd panel). On the
other hand, because of the outward angular momentum transfer, the
material in the outer part of the NS forms a disk with the maximum
density $\sim 10^{12}~{\rm g/cm^3}$ (4th panel).

\begin{figure}[thb]
\vspace{-4mm}
\begin{center}
\epsfxsize=2.7in
\leavevmode
\epsffile{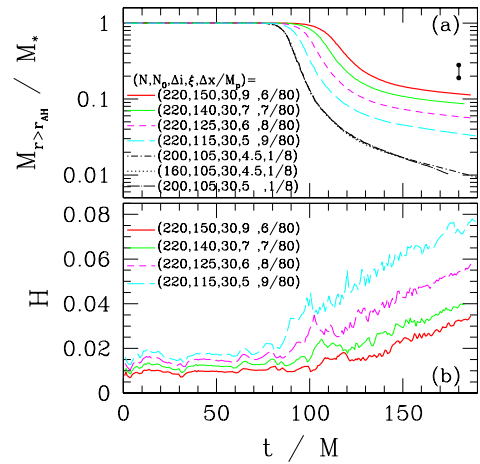}
\end{center}
\vspace{-8mm}
\caption{(a) Evolution of baryon rest-mass located outside the apparent
horizon for various grid settings. The plots for $\Delta x=M_{\rm
P}/8$ almost coincide and show that the results depend weakly on the
values of $L$ and $\xi$. On the other hand, the results depend
systematically on $\Delta x$ (see text). (b) Evolution of
averaged violation of Hamiltonian constraint for (iv)--(vii). 
\label{FIG2}}
\end{figure}

However, mass of the disk is not large. Figure 2(a) shows evolution of 
baryon rest-mass located outside apparent horizons $M_{r>r_{\rm AH}}$.
We find that $\sim 80\%$ of the mass is swallowed into the BH in $t
\sim 130M \sim 2$ ms, and swallowing continues after this time \cite{foot}. 
The value of $M_{r > r_{\rm AH}}$ depends systematically on
$\Delta x$; we find that the results for (v)--(vii) at late times
approximately obey a relation of convergence, i.e., $M_{r > r_{\rm
AH}}(t)=a(t)+b(t)\Delta x^n$ where $a(t)$ and $b(t)$ are functions of
time. The order of convergence, denoted by $n$, is between 1st and 2nd
order (i.e., $1 < n < 2$). Least-square fitting gives $a(t)$ at
$t=180M$ as $\approx 0.19M_*$ if we set $n=2$ and as $0.28M_*$ for
$n=1$ (see the solid circles in Fig. 2(a)).  Thus, the true result
should be between $0.19M_*$ and $0.28M_*$.

The adopted NS has corotating velocity field and furthermore its
radius is larger than canonical values. In reality, the disk mass
would be smaller than this value. Hence it is unlikely that a massive
disk with $\sim M_{\odot}$ is formed after merger of nonspinning BH of
mass $M > 3M_{\odot}$ and canonical NS of mass $\approx 1.4M_{\odot}$
and radius $\approx 11$--13 km, although a disk of mass of $\sim 
0.2$--$0.3M_{\odot}$ may be formed for small BH mass and short GRBs of
total energy $\sim 10^{49}$ ergs may be explained (e.g., \cite{RJ}).

Figure 2(b) shows the evolution of averaged violation of the
Hamiltonian constraint. For the average, rest-mass density is used as
a weight (see \cite{STU0}) and the integral is performed for the
region outside apparent horizons. Fig. 2(b) shows that the Hamiltonian
constraint converges approximately at 2nd order. This result is
consistent with the fact that the region except for the vicinity of
BH is followed with 2nd-order accuracy.

To summarize, we have presented our first numerical results of fully
GR simulation for merger of BH-NS binary, focusing on the case that
the BH is not spinning initially and the mass ratio $q$ is fairly
large as 0.4. It is found that even with such high value of $q$, the
NS is tidally disrupted only for the orbit very close to ISCO and 
80--90\% of the mass element is quickly swallowed into the BH
without forming massive disks.  The results do not agree
quantitatively with the prediction by the tidal approximation
study. The reasons are: (1) In the tidal approximation, one describes
NSs by the Newtonian gravity. In general relativity, gravity is
stronger and the tidal disruption is less likely. (2) The time scale
for angular momentum transfer during tidal disruption near the ISCO is
nearly as long as the plunging time scale determined by gravitational
radiation reaction and attractive force between two objects. Hence
before the tidal disruption completes, most of the material is
swallowed.

If the BH has a large spin, the final fate may be largely changed
because of the presence of spin-orbit repulsive force. This force can
weaken the attractive force between BH and NS and slow down the
orbital velocity, resulting in smaller gravitational wave luminosity
and longer radiation reaction time scale \cite{KWW}.  This effect may
help massive disk formation. The study of spinning BH binaries is one
of the next issues. The fate will also depend on EOS of NS \cite{ISM}
and mass of BH. Simulation with various EOSs and BH mass is also the
next issue.


\noindent
\underline{{\em Acknowledgements}}: 
MS thanks Y. Sekiguchi for useful discussion. 
Numerical computations were performed on the FACOM-VPP5000 at ADAC at
NAOJ and on the NEC-SX8 at YITP in Kyoto University. This work was 
supported in part by Monbukagakusho Grants (No. 17540232)
and by NSF grants PHY0071044, 0503366.


\begin{thebibliography}{99}
\bibitem{foot} $M_{r > r_{\rm AH}}$ is rest-mass of baryon located
outside apparent horizons. Mass of disk located outside an ISCO around
the formed BH is smaller than it. We follow the evolution of the
rest-mass of baryon for $r >3M$ and $r > 4.5M$ where $r=3M$--$4.5M$
are approximate locations of the ISCO and find that their values are
smaller than $M_{r > r_{\rm AH}}$ by $\sim 10\%$ and 20\%,
respectively. Thus, the disk mass would be 0.8--$0.9M_{r > r_{\rm
AH}}$ in reality.
\end{thebibliography}
\end{document}